\documentclass[5p]{elsarticle} % for two-column

\usepackage{amsmath, amssymb}
\usepackage{graphicx}

\newcommand{\rmi}{\mathrm{i}}
\newcommand{\rme}{\mathrm{e}}
\newcommand{\rmd}{\mathrm{d}}
\newcommand{\tr}{\mathop{\mathrm{tr}}\nolimits}
\begin{document}
\begin{frontmatter}
\title{Electron bunching in triple quantum dot interferometers%
}

\author{Fernando Dom\'inguez}
\author{Gloria Platero}
\author{Sigmund Kohler}
\address{Instituto de Ciencia de Materiales de Madrid, CSIC,
	Cantoblanco, E-28049 Madrid, Spain}

\date{\today}

\begin{abstract}

We study electron transport through a triple quantum dot in ring or
interferometer configuration.  In particular, we analyze the influence
of a gate voltage that detunes one of the dots, such that it becomes
off-resonant.  In this regime, interference effects fade away, i.e.,
the current becomes independent of a penetrating flux.  Despite the
absence of interference effects, the off-resonant dot causes
intriguing noise properties which we characterize by the full-counting
statistics of the transported electrons.  It turns out that the
detuning causes strong electron bunching.  Analytical results for
limiting cases support this picture.  A possible application is the
construction of current sources with widely tunable noise properties.

\end{abstract}


\end{frontmatter}

\section{Introduction}

During the last decade, a huge effort has been made to understand the
conduction properties of quantum systems that consist of only a few
discrete levels.  Spurred by Feynman's vision of ``plenty of room at
the bottom'' \cite{Feynman1960a}, it has e.g.\ been proposed to use
single molecules as elements of future electronic circuits
\cite{Ellenbogen2000a}.  Although this task is far from being
accomplished, molecular electronics already became an established
field \cite{Hanggi2002article, Cuniberti2005a}.  Apart from their
technological promises, conducting molecules may also serve as tools
for the implementation of fundamental physical phenomena.  One example
is the intriguing effect of tunnel suppression by the purely coherent
influence of an ac field \cite{Grossmann1991a}, which is stable even in
the presence of Coulomb repulsion \cite{Creffield2002a}.  Coherent
destruction of tunneling leaves its fingerprints in the transport
characteristics of laser-driven molecular wires, where an ac field
may suppress the current and its fluctuations \cite{Lehmann2003a,
Camalet2003a, Platero2004a, Kohler2005a}.

From a theoretical perspective, molecular wires share many features
with quantum dots.  In particular, the electronic structure of both
systems consists of discrete states.  Owing to this fact, coherently
coupled quantum dots may be considered as ``artificial molecules'',
despite the fact that their energy scales are several orders of
magnitude smaller than those of real molecules.  For roughly one
decade, the state of the art has been to couple just two quantum dots
coherently \cite{VanderVaart1995a, Blick1996a, vanderWiel2003a}, while
triple quantum dots have been realized only recently
\cite{Schroer2007a, Gaudreau2006a, Rogge2008a}.
Both double quantum dots \cite{Gustavsson2007a} and triple quantum
dots \cite{Gaudreau2006a, Rogge2008a} can be constructed such that
electrons coming from the source may proceed on two different paths
towards the drain.  There they interfere constructively or
destructively, depending on the setup and a possible flux enclosed by
the interfering paths.  Moreover, there exist dark states which are
superpositions of states that are decoupled from the drain and, thus,
block the electron transport \cite{Michaelis2006a, Poltl2009a}.  Such
blockade may be resolved by excitations with proper ac fields
\cite{Sanchez2006a, Busl2010a} which also create spin correlations
between transported electrons \cite{Sanchez2008c}.

\begin{figure}[tb]
\begin{center}
\includegraphics{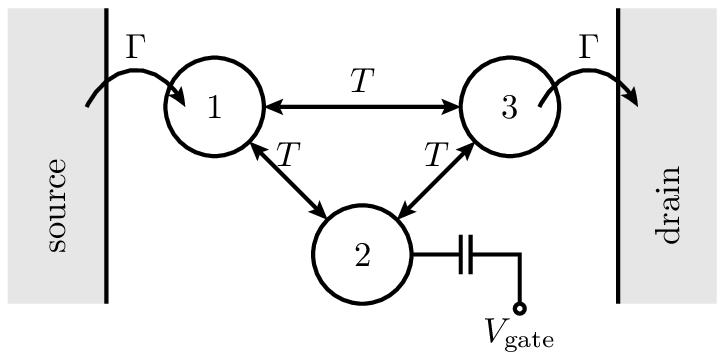}
\end{center}
\caption {\label{fig:setup}
Triple quantum dot in ring configuration coupled to electron source
and drain.  The onsite energies of dots 1 and 3 are $\epsilon_{1,3} =
0$, while the onsite energy of dot~2 can be tuned by a gate voltage
such that $\epsilon_2 = eV_{\rm gate}$.  The bias voltage $V$ is
assumed to shift the chemical potentials symmetrically, such that
$\mu_\mathrm{L} = eV/2 = -\mu_\mathrm{R}$}
\end{figure}

Here we investigate the conduction properties of a triple quantum dot
in ring configuration as sketched in Fig.~\ref{fig:setup}.  A
penetrating flux allows the operation as Aharonov-Bohm interferometer.
We focus on the influence of a gate voltage that allows shifting of
the levels of dot~2 out of resonance.  The intuitive expectation is
that for strong detuning, the two other dots govern the transport
process, such that the triple quantum dot behaves like a double
quantum dot.  While this is indeed the case for the current, electrons
temporarily trapped in dot~2 may cause Coulomb blockade and, thus,
interrupt the transport such that the electron flow becomes
avalanche-like.  This is reflected in the zero-frequency component of
the current-current correlation function which significantly exceeds
the value for a Poisson process.  With this criterion, avalanche-like
transport has been predicted for quantum dots coupled to a harmonic
mode \cite{Flindt2004a, Koch2005a}, multi-level quantum dots
\cite{Belzig2005a}, triple quantum dots in the Kondo regime
\cite{Vernek2009a}, and also when destructive interference effects
strongly suppress the current \cite{Urban2008a, Urban2009a, Li2009a,
Schaller2009a}.

An established theoretical tool for characterizing such correlated
transport is full counting statistics \cite{Belzig2005a, Lesovik1994a,
Bagrets2003a} which provides the complete information about the
distribution of the transported charge in terms of the corresponding
cumulants.  For a Markovian master equation as employed below, it is
possible to express these cumulants as derivatives of a particular
eigenvalue of the Liouville operator augmented by a counting variable
\cite{Bagrets2003a}.  Generally, this requires the computation of
derivatives of high order.  As soon as one has to rely on a numerical
treatment, practical calculations may represent a formidable task.
Recently, Flindt \textit{et al.} have found a way to circumvent this
difficulty \cite{Flindt2008a, Flindt2010a}.  Based on
Rayleigh-Schr\"odinger perturbation theory, they derived a scheme that
allows one to recursively compute the full counting statistics.  We
use this approach for our numerical solution.

Our paper is organized as follows. In Section~\ref{sec:model} we
introduce a system-lead Hamiltonian and derive a master equation
formalism for the computation of the full counting statistics.  The
numerical results for the current and the zero-frequency noise,
presented in Section~\ref{sec:numerics}, provide information about the
super-Poissonian nature of the transport process.  In
Section~\ref{sec:limits}, we derive for the cumulants up to fourth
order analytical expressions valid in the low-bias regime.
%---------------------

%--------------------------------------------------------------------
\section{Model and master equation approach}
\label{sec:model}

\subsection{Dot-lead Hamiltonian}

We consider a triple quantum dot with a ring-shaped geometry as
depicted in Fig.~\ref{fig:setup}, which is described by the Hamiltonian
\begin{equation}
\begin{split}
H_\text{TQD}
=& \sum_{i}\epsilon_i c^{\dagger}_ic_i
   + T\sum_{i,j<i} (c^{\dagger}_ic_j + c_jc^{\dagger}_i)
%\\ &
+ U\sum_{i,j<i}c^{\dagger}_ic_ic^{\dagger}_jc_j,
\end{split}
\end{equation}
where $i,j=1,2,3$ label the dots, while $c_i^{\dagger}$ and $c_i$
denote the corresponding
creation and annihilation operator of a spinless electron.
The first term refers to the onsite energy $\epsilon_i$ of an electron
on dot $i$, for which we assume that $\epsilon_1=\epsilon_3=0$, while
$\epsilon_2 = eV_\text{gate}$ can be tuned by a gate voltage.  The
second term describes electron tunneling between the dots, for which
we assume that the tunnel matrix element $T$ is the same for all three
possible transitions.
The last contribution models inter-dot Coulomb repulsion with strength
$U$.  In the present case, this term can be written in the form
$\frac{1}{2}UN(N-1)$, where $N$ denotes the total number of electrons
on the three dots.  We assume that the intra-dot interaction is
practically infinite, which corresponds to modeling each dot by a
single level.

In a ring-shaped geometry, the electrons possess two paths for
traveling from the source to the drain.  Thus, interference plays a
role.  This interference can be controlled by a magnetic flux $\Phi$
penetrating the ring.  We capture this by the replacement $c_1^\dagger
c_3 \to c_1^\dagger c_3 \exp(2\pi\Phi/\Phi_0)$, where $\Phi_0 = h/e$
denotes the flux quantum.  Note that since we consider only spinless
electrons, the effect of an associated Zeeman splitting is ignored.
Nevertheless, the flux dependence of the current provides information
about the relevance of interference effects.

Dots 1 and 3 are attached to electron reservoirs with chemical
potentials $\mu_\mathrm{L}$ and $\mu_\mathrm{R}$ such that the bias
voltage fulfills $eV = \mu_\mathrm{L}-\mu_\mathrm{R}$.  They are
described by the lead Hamiltonian
\begin{equation}
H_\text{leads}
=\sum_{\ell,q} \epsilon_{q} c_{\ell q}^\dag c_{\ell q} ,
\end{equation}
where $\ell=\mathrm{L},\mathrm{R}$, and the expectation value $\langle
c_{\ell q}^\dag c_{\ell'q'}\rangle = f(\epsilon_q-\mu_\ell)
\delta_{\ell\ell'} \delta_{qq'} \equiv f_\ell(\epsilon_q)
\delta_{\ell\ell'} \delta_{qq'} $ with chemical potential $\mu_\ell$
and the Fermi function $f(x) = [\exp(x/k_BT)+1]^{-1}$.  The dot-lead
contact is established by the tunnel Hamiltonian
\begin{equation}
\label{Hdotlead}
H_\text{tun} = \sum_{q} (V_{\mathrm{L} q} c_{\mathrm{L} q}^\dag c_1
               + V_{\mathrm{R} q} c_{\mathrm{R} q}^\dag c_3 ) + \text{h.c.}
\end{equation}
We assume within a wide-band limit that all effective coupling
strengths $\Gamma_\ell(\epsilon) = 2\pi\sum_q |V_{\ell q}|^2
\delta(\epsilon-\epsilon_q)$ are energy independent and that the setup
is symmetric such that $\Gamma_\mathrm{L} = \Gamma_\mathrm{R} = \Gamma$.

\subsection{Markovian master equation}

The derivation of a master equation starts from the Liouville-von
Neumann equation for the full density operator, $\rmi\hbar\dot R =
[H_\text{TQD}+H_\text{tun}+H_\text{leads},R]$.  By standard techniques
\cite{Blum1996a}, one obtains for the reduced density operator of the
central system, $\rho = \tr_\text{leads}R$, within second-order
perturbation theory the Bloch-Redfield equation
\begin{align}
\label{ME-rho}
\dot \rho
&= -\frac{\rmi}{\hbar}[ H_\text{TQD}, \rho] -\frac{1}{\hbar^2} \tr_\text{leads}
   \int_0^\infty \rmd\tau [H_\text{tun},[\tilde H_\text{tun}(-\tau), R]]\nonumber
\\
&\equiv \mathcal{L}\rho ,
\end{align}
which can be evaluated under the factorization assumption $R =
\rho_{\text{leads},0} \otimes \rho$.  The tilde denotes the
interaction picture with respect to the uncoupled Hamiltonian,
$\tilde X(t) = U_0^\dagger(t) X U_0(t)$, where
$U_0(t) = \exp\{-\rmi(H_\text{TQD}+H_\text{leads})t/\hbar\}$.
We proceed by inserting the dot-lead tunnel Hamiltonian
\eqref{Hdotlead} and evaluate the trace over the lead states.  In
order to cope with the interaction picture, we decompose the resulting
master equation into the eigenstates $|\alpha\rangle$ of
$H_\mathrm{TQD}$, where $E_\alpha$ denotes the corresponding
eigenenergy.  This allows us to evaluate the $\tau$-integration, which
yields a delta function and a principal value term.  Neglecting the
latter, we obtain for the density matrix elements the equation of
motion
\begin{equation}
\label{ME}
\dot\rho_{\alpha\beta}
= -\frac{\rmi}{\hbar}(E_\alpha-E_\beta)\rho_{\alpha\beta}
  + \sum_{\alpha',\beta'} \mathcal{L}_{\alpha\beta,\alpha'\beta'}
    \rho_{\alpha'\beta'}
\end{equation}
with the incoherent dot-lead tunneling given by
\begin{equation}
\begin{split}
\mathcal{L}_{\alpha\beta,\alpha'\beta'} \qquad &
\\
= \sum_{\ell=1,3} \frac{\Gamma_\ell}{2}\Big\{
&
\langle\alpha|c_\ell^\dag|\alpha'\rangle\langle\beta'|c_\ell|\beta\rangle
    f_\ell(E_\alpha{-}E_{\alpha'})
\\
+&
\langle\alpha|c_\ell^\dag|\alpha'\rangle\langle\beta'|c_\ell|\beta\rangle
    f_\ell(E_\beta{-}E_{\beta'})
\\
-& \sum_{\gamma}
   \langle\beta'|c_\ell|\gamma\rangle\langle\gamma|c_\ell^\dag|\beta\rangle
   f_\ell(E_{\gamma}{-}E_{\beta'}) \delta_{\alpha\alpha'}
\\
-& \sum_{\gamma}
   \langle\alpha|c_\ell|\gamma\rangle\langle\gamma|c_\ell^\dag|\alpha'\rangle
   f_\ell(E_{\gamma}{-}E_{\alpha'}) \delta_{\beta\beta'}
\Big\}
\\  + (c_\ell,c_\ell^\dag,  &  f_\ell)
   \to (c_\ell^\dag,c_\ell,1-f_\ell) ,
\end{split}
\end{equation}
where the leads now are labeled by the number $\ell$ of the dot to
which they are attached.  Electron tunneling from the leads to the dots
is described by the explicitly written terms, while the replacement in
the last line yields the terms for tunneling from the dots to the
leads.

The solution of the master equation \eqref{ME} contains the full
information about the state of the central system and provides all
corresponding expectation values.  However, we are interested in the
statistics of the charge transported after the initial preparation,
which is an expectation value of lead operators.  Therefore we have to
generalize the master equation formalism by introducing a counting
variable which allows one to keep track of this information.

%-----------------------------------------------------------------------------
\subsection{Full-counting statistics}

The counting variable $\chi$ is defined via the moment
generating function
\begin{equation}
\phi(\chi,t) = \langle\exp(\rmi\chi N_\mathrm{R})\rangle_t \,,
\end{equation}
where $N_\mathrm{R} = \sum_q c_{3q}^\dagger c_{3q}$ is the electron
number operator of the right lead, while the angular brackets refer to the
expectation value at time $t$.  The $k$th derivative of $\phi(\chi,t)$ with
respect to $\rmi\chi$ at $\chi=0$ obviously is the moment $\langle
N_\mathrm{R}^k\rangle$ of the electron distribution in the right lead.
The cumulants of the distribution are defined as the corresponding
derivatives of $\ln\phi(\chi,t)$.  For a Markov process, they
eventually become linear in time. Their time-derivatives at large
times,
\begin{equation}
\label{ck}
C_k = \frac{\partial}{\partial t} \frac{\partial^k}{\partial(\rmi\chi)^k}
      \ln\phi(\chi,t)\Big|_{\chi=0, t\to\infty}
\end{equation}
are the stationary current cumulants and characterize the transport.
The first and the second cumulant, $C_1$ and $C_2$, are essentially
the current $I = eC_1$ and its zero-frequency noise $S=e^2C_2$,
respectively.  Their ratio, the Fano factor $F=C_2/C_1$, represents a
dimensionless measure for the noise of the transport process
\cite{Fano1947a}.  It is defined such that for a Poissonian process
$F=1$.

Our goal is now to find a reduced master equation that allows one to
compute $\phi(\chi,t)$.  We start again from the full density operator
$R$, but now multiply it with the operator $\exp(\rmi\chi
N_\mathrm{R})$ before tracing out the leads.  This yields the
generalized reduced density operator $P(\chi,t) = \tr_\text{leads}\{
\exp(\rmi\chi N_\mathrm{R})R\}$ which contains information about the
electron distribution of the drain and fulfills the trace condition
$\tr P = \phi$.  Using the commutation relations
$[N_\mathrm{R},\Lambda] = \Lambda$ and $[N_\mathrm{R},\Lambda^\dagger]
= -\Lambda^\dagger$, where $\Lambda = \sum_q V_q c_3^\dagger c_{3q}$,
we obtain the master equation
\begin{equation}
\label{EOM-P}
\dot P(\chi,t) = \mathcal{L}_\chi P(\chi,t)
\end{equation}
with the augmented Liouvillian
\begin{equation}
\label{Lchi}
\mathcal{L}_\chi = \mathcal{L}
+ (\rme^{\rmi\chi}-1)\mathcal{J}^+ + (\rme^{-\rmi\chi}-1)\mathcal{J}^-
\end{equation}
and the particle current superoperators $\mathcal{J}^\pm$. After some
algebra, we find for the latter the expression \cite{Kaiser2007a}
\begin{equation}
\label{eq:Jotamas}
\mathcal{J}^- \rho
=\frac{e\Gamma_\ell}{2\pi}\int_0^\infty \rmd\tau\int \rmd\epsilon\,
  \rme^{-\rmi\epsilon\tau}\tilde{c}_3^\dag(-\tau)
  \rho c_3 f_\ell(\epsilon) +\text{h.c.},
\end{equation}
and $\mathcal{J}^+$ is formally obtained from
$\mathcal{J}^-$ by the replacement
$(c_3^\dagger,c_3,f_\mathrm{R}) \to (c_3,c_3^\dagger,
1-f_\mathrm{R})$.  The superoperator $\mathcal{J}^+$ describes
tunneling from dot 3 to the right lead, while $\mathcal{J}^-$
corresponds to the reversed process.
We proceed as for the derivation of the master equation \eqref{ME} and
decompose the current superoperators into the eigenstates of
$H_\mathrm{TQD}$ which yields
\begin{align}
\mathcal{J}^+_{\alpha\beta,\alpha'\beta'}
= &
  \frac{\Gamma_\mathrm{R}}{2}
  \langle\alpha|c_3|\alpha'\rangle \langle\alpha'|c_3^\dag|\beta\rangle
\\ & \times \big\{
        2- f_\mathrm{R}(E_\alpha{-}E_{\alpha'})
         - f_\mathrm{R}(E_\beta{-}E_{\beta'})  \big\},
  \nonumber
\\
\mathcal{J}^-_{\alpha\beta,\alpha'\beta'}
= &
  \frac{\Gamma_\mathrm{R}}{2}
  \langle\alpha|c_3^\dag|\alpha'\rangle \langle\alpha'|c_3|\beta\rangle
\\ & \times \big\{
           f_\mathrm{R}(E_\alpha{-}E_{\alpha'})
         + f_\mathrm{R}(E_\beta{-}E_{\beta'})  \big\} .
  \nonumber
\end{align}

In the long-time limit, the dynamics of $P(\chi,t)$ is governed by the
eigenvalue of $\mathcal{L}_\chi$ with the largest real part, denoted
as $\lambda(\chi)$.  We assume that $\lambda(\chi)$ can be uniquely
identified by its limit $\lim_{\chi\to 0}\lambda(\chi)=0$, i.e., it
corresponds to the stationary solution of the Liouvillian $\mathcal{L}
= \mathcal{L}_{\chi\to 0}$.  Then, $P = A(\chi) \exp[\lambda(\chi)
t]$, with $A(\chi)$ be the corresponding ``eigenoperator'' of
$\mathcal{L}_\chi$.  It is straightforward to see that
$\ln\phi(\chi,t) = \ln\tr A(\chi) + \lambda(\chi)t$.  In the long-time
limit, the contribution of the normalization factor is not relevant
and, thus, we can conclude that $\lambda(\chi)$ is the current
cumulant generating function \cite{Bagrets2003a}.  It can be written
as the series
\begin{equation}
\label{lambda-series}
\lambda(\chi) = \sum_{k=1}^\infty \frac{C_k}{k!} (\rmi\chi)^k .
\end{equation}
The remaining task is now to compute the proper eigenvalue of
$\mathcal{L}_\chi$ and its derivatives with respect to $\chi$.  In
many cases, the reduced density matrix has a sufficiently small
dimension or possesses symmetries, such that one can continue with
analytical calculations.
As soon as one has to resort to a numerical treatment, however, one
faces the difficulty of numerically computing derivatives.  This can
be avoided with the recursive scheme developed in
Ref.~\cite{Flindt2008a} even for non-Markovian master
equations.  Here we restrict ourselves to the Markovian limit.

Since we are interested in the derivatives of the cumulant generating
function at $\chi=0$, we can treat $\chi$ as small parameter and
employ perturbation theory.  Then the series \eqref{lambda-series} for
the eigenvalue $\lambda(\chi)$ corresponds to the usual ansatz for the
eigenvalue. It can be computed by the recursion derived in
Appendix~\ref{app:perturbation}, see Eqs.~\eqref{app:iterE} and
\eqref{app:iterPhi}.  Upon setting in these expressions $E_k =
C_k/k!$,  $V_k = [\mathcal{J}^++(-1)^k\mathcal{J}^-]/k!$, and $P_k =
|\phi_k\rangle/k!$, we find
\begin{align}
C_k &= \sum_{k'=0}^{k-1} \begin{pmatrix}k\\k'\end{pmatrix}
\tr \{\mathcal{J}^+ +(-1)^{k-k'}\mathcal{J}^-\} P_{k'} ,
\\
P_k &= \frac{\mathcal{Q}}{\mathcal{L}} \sum_{k'=0}^{k-1}
\begin{pmatrix}k\\k'\end{pmatrix}
\Big\{ C_{k-k'} - [\mathcal{J}^++(-1)^{k-k'}\mathcal{J}^-]
\Big\} P_{k'} .
\end{align}
The iteration starts from the stationary solution of the Liouvillian,
$P_0 = |\phi_0\rangle = \rho_\infty$, which implies $E_0 = C_0 =0$.
Multiplication with the corresponding left eigenvector
$\langle\phi_0|$ must correspond to computing the trace.  This becomes
clear when one notes that the Liouvillian is trace conserving, i.e.\
$\tr\mathcal{L}X=0$ for any operator $X$.  Consequently, $\mathcal{Q}
= (\mathbf{1}-\rho_\infty\tr)$ is the projector on the subspace
perpendicular to $\rho_\infty$.  In this subspace, the Liouvillian
possesses the pseudo-inverse $\mathcal{Q}/\mathcal{L}$.  Applying it
to any operator, i.e.\ computing $(\mathcal{Q}/\mathcal{L})B \equiv
X$, is equivalent to solving the linear equation $\mathcal{L}X = B$
under the constraint $\tr X= 0$.

In particular, we obtain for the first two cumulants, which are the
current and the zero-frequency noise, the known expressions
\cite{Flindt2004a}
\begin{align}
I ={} & e\tr (\mathcal{J}^+-\mathcal{J}^-)\rho_\infty
\label{eq:theocurrent}
\\
S ={} & e^2\tr (\mathcal{J}^++\mathcal{J}^-)\rho_\infty
\nonumber \\ &
    - 2e^2\tr(\mathcal{J}^+-\mathcal{J}^-) \frac{\mathcal{Q}}{\mathcal{L}}
      (\mathcal{J}^+-\mathcal{J}^-)\rho_\infty.
\label{eq:theonoise}
\end{align}

%---------------------------------------------------------------------
\section{Numerical observations}
\label{sec:numerics}

An intuitive picture for the transport properties as a function of the
gate voltage can be provided for the limiting cases in which the
detuning of dot~2 is either much larger or much smaller than the
inter-dot tunneling:  In the limit $|eV_\text{gate}| \ll T$, all three
dots are in resonance and, thus, the electrons may take with similar
probability two different routes, namely $|1\rangle \rightarrow
|3\rangle$ and $|1\rangle \rightarrow |2\rangle\rightarrow |3\rangle$.
Therefore one expects interference to be relevant, such that the
current can be modified by a magnetic flux penetrating the ring.  If
the gate voltage is large, by contrast, i.e.\ for $|eV_\text{gate}|
\gg T$, dot~2 is off-resonant and, thus, should be of minor relevance.
Therefore, the transport properties of the setup are expected to be
essentially those of a double quantum dot formed by dots 1 and 3.  Our
numerical results, however, will demonstrate that this is only true
for the average current, while the current noise is significantly
altered by the presence of dot~2 even when far detuned.

As a first step, we solve the master equation numerically.  In doing
so, we identify parameter regimes in which the stationary state of the
triple dot is dominated by a few eigenstates.  This allows us to
reduce the complexity of the master equation, such that we can achieve
an analytical treatment that provides insight to the transport
mechanism.
For the numerical solution, we use the dot-lead coupling $\Gamma$ as
energy unit.  For the typical value $\Gamma = 10\mu\mathrm{eV}$, the
corresponding current unit reads $e\Gamma/\hbar = 2.43\mathrm{nA}$.

\subsection{Stationary current and occupation probabilities}

\begin{figure}[tb]
\begin{center}
\includegraphics{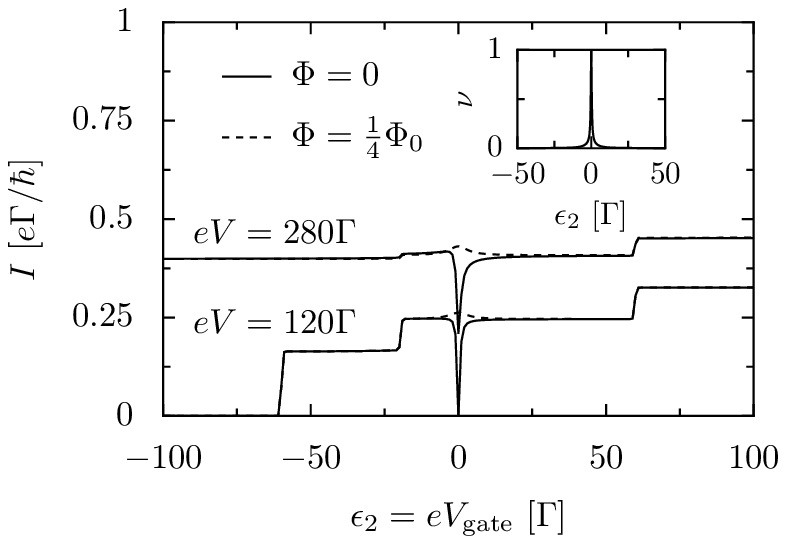}
\end{center}
\caption{\label{fig:I_epsilon}
Current as a function of the gate voltage $V_\text{gate} =
\epsilon_2/e$ for different magnetic fluxes $\Phi$ and bias voltages.
The inter-dot tunnel coupling and interaction are $T=2\Gamma$
and $U = 80\Gamma$, respectively.
Inset: Corresponding visibility for $eV=120\Gamma$.}
\end{figure}

The stationary current as a function of the gate voltage
$V_\text{gate} = \epsilon_2/e$ is shown in Fig.~\ref{fig:I_epsilon}.  It
demonstrates that for $|\epsilon_2| \lesssim T$, the current indeed is
significantly flux dependent as conjectured above.  For flux $\Phi=0$,
the current assumes its minimum.  This can be explained by the fact
that for $\Phi=0=\epsilon_2$, the state $|\psi_\text{dark}\rangle =
(|1\rangle-|2\rangle)/\sqrt{2}$ is an eigenstate of the triple dot
Hamiltonian and is orthogonal to state $|3\rangle$.  Therefore it
is not directly coupled to the right lead, which has the
consequence that $|\psi_\text{dark}\rangle$ may trap an electron which
henceforth blocks transport \cite{Emary2007b}.  This is formally
related to an atomic state that is decoupled from a light field and,
thus, remains invisible \cite{Arimondo1996a}.
%--------------------
%--------------------

As a criterion for the relevance of interference, we compute the
visibility
\begin{align}
\nu=\frac{I_\text{max}-I_\text{min}}{I_\text{max}+I_\text{min}},
\label{visibility}
\end{align}
where $I_\text{max} = \mathop{\mathrm{max}}_\Phi I(\Phi)$ is the
maximal current upon flux variation for a given gate voltage, and
$I_\text{min}$ is defined accordingly.  The inset of
Fig.~\ref{fig:I_epsilon} demonstrates that for $|\epsilon_2| \gg T$,
interference plays a minor role.  The reason for this is that when
dot~2 is strongly detuned, transport through this off-resonant dot
requires co-tunneling and, thus, the path via dot~2 has a
significantly lower probability than the direct path.

\begin{figure}[tb]
\begin{center}
\includegraphics{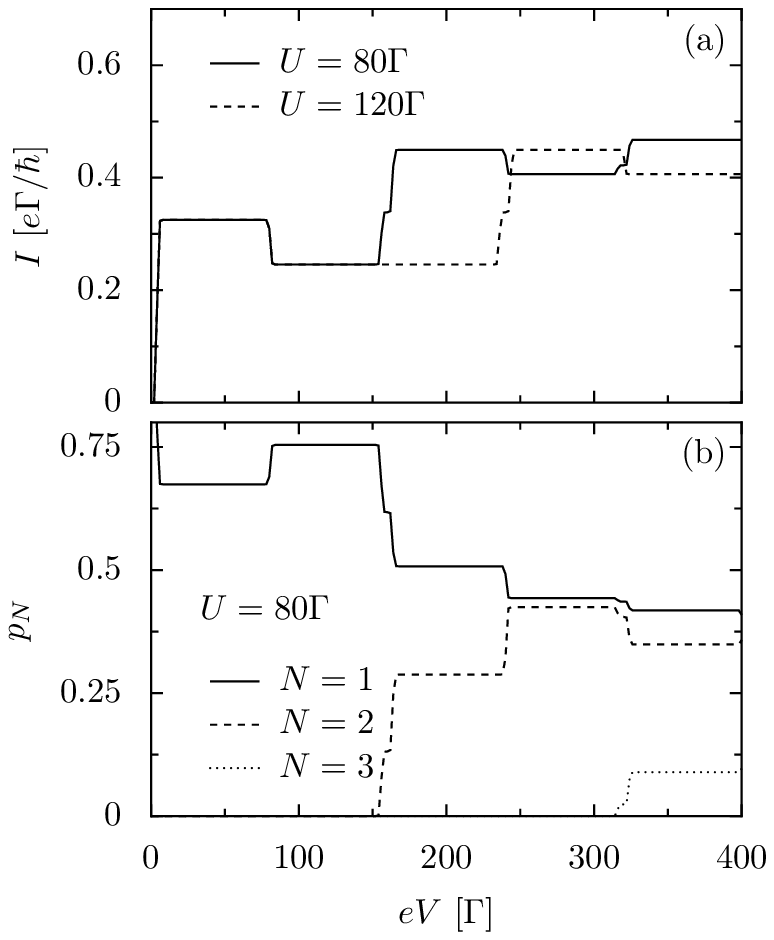}
\end{center}
\caption{\label{fig:current}
(a) Current as a function of the bias voltage $eV = \mu_\mathrm{L} -
\mu_\mathrm{R}$ for various interaction strengths.  The inter-dot
tunnel coupling and the gate voltage are $T=2\Gamma$ and
$V_\text{gate} = \epsilon_2/e = 40\Gamma/e$, respectively.
(b) Probability that for $U=80\Gamma$, the asymptotic state of the
triple quantum dot contains $N$ electrons.}
\end{figure}

Henceforth we focus on the region \textit{without} interference
effects.  There a main feature of the current is that it exhibits
plateaus, see Fig.~\ref{fig:current}(a).  This is consistent with the
usual Coulomb blockade scenario in which the bias and the gate
voltage, determine the maximal number of electrons that can reside on the
dots.  Accordingly, the probability for having a particular dot
occupation changes at the steps, as can be appreciated from
Fig.~\ref{fig:current}(b).  Moreover, an electron can tunnel from a
left lead to the dots only if it has sufficient energy to compensate the
Coulomb repulsion of the electrons that are already in the triple
quantum dot.  Thus the occupation with $N$ electrons ($N=1,2,3$) is
possible only if one chemical potential is larger than $(N-1)U$.  For
the symmetric positive voltage drop assumed herein, the larger
chemical potential is $\mu_\mathrm{L} = eV/2$.  This implies that the
occupation probability $p_N$ vanishes if $eV<2(N-1)U$, which is consistent
with the occurrence of the steps shown in Fig.~\ref{fig:current}(b).
In particular, for $e|V|<2U$, only single occupation plays a role.
Below we will use this fact for establishing an analytical treatment
in the low-bias regime.
%--------------------
%--------------------

\subsection{Shot noise and Fano factor}

\begin{figure}[tb]
\begin{center}
\includegraphics{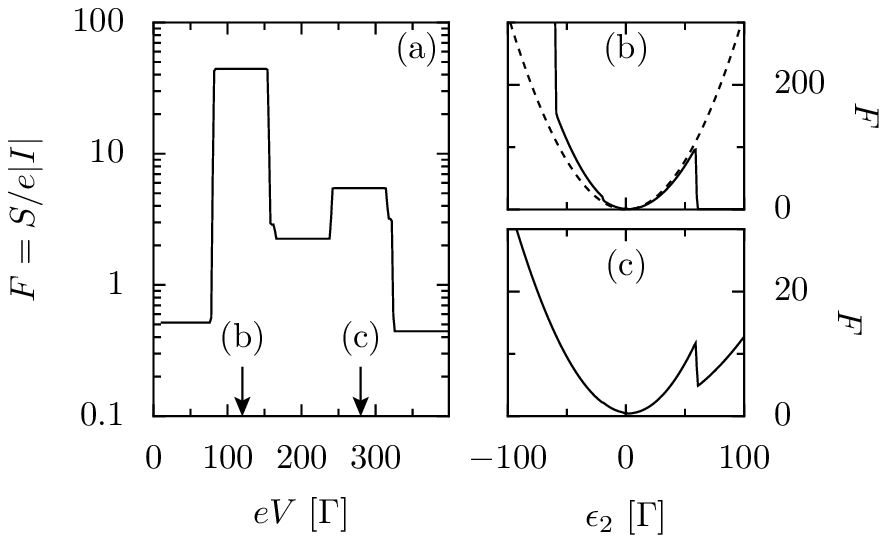}
\end{center}
\caption {\label{fig:fano}
(a) Fano factor of the current shown in Fig.~\ref{fig:current}(a) for the
interaction strength $U=80\Gamma$.
(b,c) Fano factor as a function of the detuning $\epsilon_2 =
eV_\text{gate}$ for the bias voltages (b) $V = 120\Gamma/e$ and
(c) $V=280\Gamma/e$.
The dashed line in panel (b) represents the analytical result
\eqref{F1e} valid in the one-electron regime.}
\end{figure}

So far we have seen that unless the bias voltage is extremely low, the
current is always of the order $e\Gamma/\hbar$, i.e.\ changing the
gate voltage or the bias voltage modifies the current typically by a
factor of the order unity.  Figure~\ref{fig:fano} demonstrates that
the noise, characterized by the Fano factor $F = S/e|I|$, generally
exhibits a more significant dependence on both the bias and the gate
voltage.

In the three-electron regime, i.e.\ for $eV>4U$, the Fano factor is
of order unity and almost independent of $\epsilon_2$.
This indicates that the transport process is Poissonian.  In the
two-electron regime, i.e.\ for $U < eV/2 < 2U$, the Fano factor
[Fig.~\ref{fig:fano}(c)] exhibits a quadratic dependence on
$\epsilon_2$ as well, but the absolute values are now significantly
smaller.  

A rather pronounced dependence on the gate voltage $V_\mathrm{gate} =
\epsilon_2/e$ is found in that part of the single-electron regime in
which the left chemical potential is so large that all single particle
levels lie within the voltage window, i.e.\ for $\epsilon_2<eV/2<U$;
see Fig.~\ref{fig:fano}(b).  In particular, we observe a quadratic
growth of the Fano factor, $F\propto \epsilon_2^2$, with highly
super-Poissonian values.  This already indicates electron bunching,
where each bunch consists of roughly $F$ electrons
\cite{Levitov2004a}.
%---------------
%---------------

\subsection{Full counting statistics}
\label{sec:fcs}

A more complete picture can be drawn by considering the distribution
function of the transported electrons or equivalently all cumulants,
i.e., the full counting statistics.  Here we restrict ourselves to the
corresponding long-time limit which characterizes the low-frequency
fluctuations.  For avalanche-like transport, higher-order cumulants
were computed \cite{Flindt2004a, Belzig2005a, Urban2008a,
Urban2009a, Li2009a} and recently also measured \cite{Gabelli2009a}.

In a simplified picture, one may consider the electron avalanches as
particles with charge $q = eF$ that are transported in an uncorrelated
manner.  For this Poisson process, the cumulants~\eqref{ck} grow
exponentially fast with their index according to the relation
\cite{vanKampen1992a}
\begin{equation}
\label{fcs_hypothesis}
C_k = \Big(\frac{q}{e}\Big)^{k-1} C_1 = F^{k-1}C_1.
\end{equation}
Closer theoretical investigations of avalanches with finite duration
\cite{Belzig2005a, Flindt2010a}, however, indicate that this picture
deserves some refinement.  In fact there it was found that the
cumulants grow even faster.  In avalanche diodes, by contrast, the
opposite was measured, namely that the cumulants do not grow as
fast \cite{Gabelli2009a}.  But so far the underlying mechanism has not
been revealed.

\begin{figure}[tb]
\begin{center}
\includegraphics{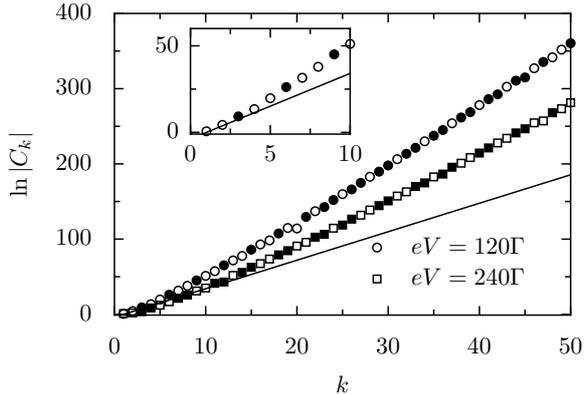}
\end{center}
\caption {\label{fig:fcs}
Full counting statistics of the current for $\epsilon_2=40\Gamma$ and
the bias voltages $V=120\Gamma/e$ (one-electron regime) and
$V=280\Gamma/e$ (two-electron regime) expressed by the cumulants
$C_k$.  All other parameters are as in Fig.~\ref{fig:fano}.  Open
symbols mark positive values while filled symbols correspond to
negative values.  The line marks the behavior for uncorrelated tunneling of
charges $q=eF$ in the one-electron regime (circles), cf.\
Eq.~\eqref{fcs_hypothesis}.
Inset: Enlargement of the lower left corner.}
\end{figure}

Figure \ref{fig:fcs} depicts the super exponential growth of the $C_k$
in both the one-electron regime and the two-electron regime.  The
solid line demonstrates that only the cumulants of very low order
follow the behavior of the Poisson process underlying
Eq.~\eqref{fcs_hypothesis}.  Thus, we also in our case observe that
already the third cumulant slightly exceeds the value $F^2C_1$, while
higher-order cumulants even assume significantly larger absolute
values.  It is also interesting to notice that the sign of the
cumulants changes periodically with the index $k$.  This hints on the
recently conjectured ``universal cumulant oscillations''
\cite{Flindt2009a}.
%-----------------

%-----------------------------------------------------------------------
\section{Analytical solution in the one-electron regime}
\label{sec:limits}

If the bias voltage is so small that only one electron can reside on
the triple dot, i.e.\ for $eV < 2U$, all relevant eigenstates of
$H_\text{TQD}$ are the empty state $|0\rangle$ and the one-electron
states.  To lowest order in $T/\epsilon_2$, the latter are given in the
basis of the localized states $|j\rangle = c_j^\dagger|0\rangle$ by
the expressions
\begin{align}
|\phi_1\rangle
={} & \frac{1}{\sqrt{2}}\left(|1\rangle-|3\rangle\right) ,
\label{1e:ef1}
\\
|\phi_2\rangle
={} & \frac{1}{\sqrt{2+4(T/\epsilon_2)^2}}
      \left(|1\rangle -\frac{2T}{\epsilon_2}|2\rangle +|3\rangle\right) ,
\label{1e:ef2}
\\
|\phi_3\rangle
={} & \frac{1}{\sqrt{1+2(T/\epsilon_2)^2}}
   \left(\frac{T}{\epsilon_2}|1\rangle +|2\rangle
   +\frac{T}{\epsilon_2}|3\rangle\right).
\label{1e:ef3}
\end{align}
For symmetry reasons, each of these states couples with equal strength
to the left and to the right lead, such that the corresponding
incoherent transition rates from and to the leads are given by
\begin{equation}
\Gamma_{\mathrm{L},n} =\Gamma_{\mathrm{R},n}
=\Gamma|\langle \phi_n|3\rangle|^2 \equiv \Gamma_n .
\label{eq:gamma}
\end{equation}
For the approximate eigenstates \eqref{1e:ef1}--\eqref{1e:ef3} the
coupling constants read $\Gamma_1 = \Gamma_2 = \Gamma/2$ and $\Gamma_3
= \Gamma T^2/\epsilon_2^2$.

If the dot-lead coupling $\Gamma$ is sufficiently small, one can
neglect within a rotating-wave approximation off-diagonal elements of
the reduced density operator \cite{Kohler2005a}, such that the system
is well described by the occupation probabilities $(P_0,P_1,P_2,P_3)$
for the eigenstates $|0\rangle$ and $|\phi_n\rangle$, $n=1,2,3$.  The
corresponding Liouville operator is
\begin{equation}
\label{L1e}
\mathcal{L}_{1e}= \begin{pmatrix}
                  -\Gamma & \Gamma_1 & \Gamma_2 & \Gamma_3\\
                  \Gamma_1 & -\Gamma_1 & 0 & 0\\
                  \Gamma_2 & 0 & -\Gamma_2 & 0\\
                  \Gamma_3 & 0 & 0 & -\Gamma_3
                  \end{pmatrix},
\end{equation}
where the index ``$1e$'' refers to the restriction to single
occupation.  We have assumed that the chemical potentials are such
that all three one-electron states lie within the voltage window.
Moreover, we have employed the sum rule $\Gamma = \sum_{n=1}^3
\Gamma_n$ which follows from the completeness relation for the
eigenstates $|\phi_n\rangle$ in the one-electron subspace.  The
corresponding particle current operators are $\mathcal{J}^-=0$, while
$\mathcal{J}^+$ is obtained from $\mathcal{L}_{1e}$ by keeping only
the elements placed above the diagonal.

The stationary state of the Liouville operator can now be readily
computed and reads $P_0 = 1/4 = P_n$ for all $n$.  Inserting this into
the current formula \eqref{eq:theocurrent}, yields $I=e\Gamma/4\hbar$.
Interestingly enough, the current does not depend on the structure of
the eigenstates, which is a consequence of the mentioned sum rule for
the $\Gamma_n$.

For the Liouville operator \eqref{L1e}, not only the stationary
current, but also the zero-frequency noise \eqref{eq:theonoise} can be
evaluated exactly.  Starting from the latter expression, one obtains
after some lines of straightforward calculation the result
\begin{equation}
S = \frac{e^2}{32\hbar}\sum_{n,n'=1}^3\frac{\Gamma_n^2}{\Gamma_{n'}}
   +\frac{e^2}{16\hbar}\sum_{n=1}^3\frac{\Gamma_1\Gamma_2\Gamma_3}{\Gamma_n^2}.
\label{eq:noise-single}
\end{equation}
In the limit $\Gamma_{1,2}\ll\Gamma_3$, only terms with $\Gamma_3$ in
the denominator are relevant, so that we obtain $S = (e^2/\hbar)
\Gamma^2/32\Gamma_3$.  With the above expressions
for $\Gamma_n$, the corresponding Fano factor becomes
\begin{equation}
\label{F1e}
F_{1e} = \frac{\epsilon_2^2}{8 T^2} .
\end{equation}
It indeed exhibits the predicted parabolic dependence on the gate
voltage $V_\text{gate} = \epsilon_2/e$.
A quantitative comparison of both the current and the Fano factor with
the numerical results depicted in Fig.~\ref{fig:fano}(b) yields a
satisfactory agreement in the range considered here.  This implies
that our one-electron model indeed captures the essential features of
the electron avalanches through the triple quantum dot.  Moreover, it
implies that the (average) size of the avalanches is
$q=e\epsilon_2^2/8T^2$.

The proportionality of the Fano factor to $(\epsilon_2/T)^2$ suggests
for the transport in the one-electron regime the following scenario:
Assume that the off-resonant dot is initially unoccupied, while
electrons are transported via dots~1 and 3.  Then, according to
standard perturbation theory, electrons in these dots may tunnel to
dot~2 with a probability amplitude proportional to the tunnel matrix
element divided by the energy difference, i.e.\ with a probability $p
\propto (T/\epsilon_2)^2$.  Thus after on average $1/p$ electrons have
been transported, an electron will tunnel to dot~2 and cause temporary
Coulomb blockade which is resolved only when the electron tunnels
further.  Thus, the dynamics is given by periods with an open channel
that on average terminate after an avalanche with $1/p \propto
(T/\epsilon)^2$ electrons has been transported. This explains the
observed proportionality of the Fano factor $F \propto
(\epsilon_2/T)^2$.  This scenario also hints on why the cumulants
deviate from conjecture \eqref{fcs_hypothesis}:  The average duration
of an avalanche is as long as the waiting time between subsequent
avalanches, while conjecture \eqref{fcs_hypothesis} is based on the
assumption of a much shorter avalanche duration.

In the limit $\Gamma_3 \ll \Gamma$, it is still possible to evaluate
some cumulants of higher order analytically, although this becomes
increasingly tedious.  Nevertheless it is worth proceeding up to forth
order for which we obtain
\begin{align}
C_1 ={}& \frac{\Gamma}{4\hbar} ,
\\
C_2 ={}& \frac{\Gamma}{4\hbar} \Big(\frac{\Gamma}{8\Gamma_3}\Big) ,
\\
C_3 ={}& -\frac{3\Gamma}{4\hbar} \Big(\frac{\Gamma}{8\Gamma_3}\Big)^2 ,
\\
C_4 ={}& \frac{3\Gamma}{4\hbar} \Big(\frac{\Gamma}{8\Gamma_3}\Big)^3 .
\end{align}
Thus, the first two cumulants behave as expected for short avalanches
with charge $q = e\Gamma/8\Gamma_3$.  However, already the third and
the forth cumulant deviate from the Poissonian value by a factor
$\pm3$ in compliance with our numerical observations in
Sec.~\ref{sec:fcs}.

\section{Conclusions}

The emergence of an interference pattern with good visibility usually
requires the coherent superposition of two or more paths that are
traversed with like probabilities.  In an Aharonov-Bohm interferometer
formed by quantum dots in ring configuration, a significant detuning
of the dot that is not connected to any lead has the consequence that
transport through one arm requires co-tunneling.  Since this reduces
the transmission probability of that path, the interference pattern as
a function of a penetrating flux will fade away.  We have demonstrated
that, nevertheless, strongly detuned interferometers bear interesting
effects manifest in the noise properties of the current.  In
particular, we have shown that strong electron bunching may occur.  It
turned out that this is most pronounced in the low bias regime in
which Coulomb repulsion forbids the occupation of the triple quantum
dot by more than one electron.

The prime quantity of interest in that context is the Fano factor for
which we have predicted huge values: the relative noise strength may
exceed that of a Poisson process by several orders of magnitude.  This
has led us to the conclusion that the current consists of avalanches
with a finite duration.  The physical reason for this is that
electrons may become trapped in the off-resonant quantum dot, such
that Coulomb blockade interrupts the transport until the trapped
electron is released.  The analysis of the higher-order charge
fluctuations---the full-counting statistics---has revealed that the
cumulants grow even super exponentially with their order.

For the computation of the full-counting statistics, we have employed
the iterative scheme recently developed in Ref.~\cite{Flindt2008a}.
This has been essential for the treatment of the three-dot problem,
since the Hilbert space is already too large for a full analytical
treatment.  Nevertheless, in the most relevant regime of single
occupation, we have performed the first iteration steps analytically,
such that we have obtained expressions for the first four cumulants in
the limit of strong detuning.

In conclusion, we have studied electron interferometers in a regime
that so far has not attracted much attention, most likely due to the
lack of pronounced interference effects.  Precisely in this regime,
however, the shot noise properties are most interesting and strong
electron bunching occurs.  Thus, quantum dots in ring configuration
may serve not only for the observation of interference effects, but
also for the creation of currents with widely tunable
super-Poissonian fluctuations.

\section*{Acknowledgements}

S.K. likes to thank Peter H\"anggi for numerous inspiring discussions
during many years of fruitful collaboration.
We acknowledge helpful discussions with Christian Flindt.
This work has been supported by the Spanish Ministry of Science and
Innovation through project MAT2008-02626, a Ram\'on y Cajal fellowship
(S.K.), and a FPI grant (F.D.).

\appendix
\section{Rayleigh-Schr\"odinger perturbation theory}
\label{app:perturbation}

In most textbooks on quantum mechanics, perturbation theory is
formulated for a perturbation linear in the small parameter, while the
augmented Liouvillian $\mathcal{L}_\chi$ contains arbitrarily high
powers of $\chi$; see Eq.~\eqref{Lchi}.
In this appendix we derive a recursive scheme for the representation
of an eigenvalue as a Taylor series in a perturbation parameter.  Our
calculation is inspired by Chap.~5 of Ref.~\cite{Sakurai}, for an
alternative derivation see Ref.~\cite{Baiesi2009a}.
Despite the fact that we formulate the problem in terms of a quantum
mechanical energy eigenvalue problem, our derivation is not restricted
to Hermitian operators.

We consider a ``Hamiltonian'' $H = H_0+V(\alpha)$ for which the
perturbation $V(\alpha)$ is an analytical function of the perturbation
parameter $\alpha$ and vanishes in the limit $\alpha\to 0$.  Thus, it
can be decomposed into the series
\begin{equation}
\label{app:V}
V(\alpha) = \sum_{k=1}^\infty \alpha^k V_k .
\end{equation}
Our goal is to find a series for the eigenvalue $E(\alpha)$
of $H$ that fulfills $\lim_{\alpha\to0} E(\alpha) = E_0$, where $E_0$
is the eigenvalue of $H_0$ corresponding to a particular eigenvector
$|\phi_0\rangle$, i.e., $H_0|\phi_0\rangle = E_0|\phi_0\rangle$.  A
central assumption is that both the eigenvalue and the corresponding
eigenvector $|\phi(\alpha)\rangle$ can be decomposed into a series in
$\alpha$ as well, such that we can employ the ansatz
\begin{align}
\label{app:E(alpha)}
E(\alpha) & = E_0 + \sum_{k=1}^\infty \alpha^k E_k \,,
\\
\label{app:phi(alpha)}
|\phi(\alpha)\rangle
&= |\phi_0\rangle + \sum_{k=1}^\infty \alpha^k |\phi_k\rangle \,.
\end{align}
For the present purpose, it is sufficient to consider only the case of
non-degenerate $E_0$.  Note that for non-Hermitian $H_0$,
$\langle\phi_0|$ is generally not the Hermitian adjoint of
$|\phi_0\rangle$, but rather the corresponding left eigenvector, i.e.,
the solution of $\langle\phi_0|H_0 = E_0\langle\phi_0|$.  We choose
the normalization such that $\langle\phi_0|\phi_0\rangle=1$.

We start by writing the eigenvalue equation $H|\phi(\alpha)\rangle =
E(\alpha)|\phi(\alpha)\rangle$ in the form
\begin{equation}
\label{app:ew}
(H_0 -E_0) |\phi(\alpha)\rangle
= \{ E(\alpha) - E_0 - V(\alpha) \} |\phi(\alpha)\rangle ,
\end{equation}
which is convenient for the derivation of a formal solution and
subsequent iteration.
Since $(H_0-E_0)|\phi_0\rangle =0$, Eq.~\eqref{app:ew} defines
$|\phi(\alpha)\rangle$ only up to a component proportional to
$|\phi_0\rangle$.  Moreover, it implies that the inverse of
$H_0-E_0$ does not exist.  It is however possible to define the
pseudo-inverse $Q(H_0-E_0)^{-1}Q$ [or in short: $Q/(H_0-E_0)$], where
$Q$ is the projector on the subspace spanned by the eigenvectors of
$H_0$ with non-zero eigenvalue.  In the present case of non-degenerate
$E_0$, it reads $Q = \mathbf{1}-|\phi_0\rangle\langle\phi_0|$.

The pseudo-inverse allows one to cast Eq.~\eqref{app:ew} into the
form
\begin{equation}
\label{app:solution}
|\phi(\alpha)\rangle
= |\phi_0\rangle
  + \frac{Q}{H_0-E_0} \{ E(\alpha) -E_0 -V(\alpha) \}|\phi(\alpha)\rangle .
\end{equation}
The first term on the right-hand side can be multiplied by any factor
without violating Eq.~\eqref{app:ew}.  We have chosen
it such that $\lim_{\alpha\to0}|\phi(\alpha)\rangle = |\phi_0\rangle$.
Moreover, since $\langle\phi_0|Q=0$, the relation
$\langle\phi_0|\phi(\alpha)\rangle = \langle\phi_0|\phi_0\rangle = 1$
holds.  An important feature of Eq.~\eqref{app:solution} is that the
second term on the right-hand side is of first order in $\alpha$.
Therefore, it can be solved iteratively in the following way.

Multiplying Eq.~\eqref{app:ew} by $\langle\phi_0|$, we find that the
left-hand-side vanishes owing to $\langle\phi_0|H_0 =
E_0\langle\phi_0|$.  Thus we obtain $E(\alpha)-E_0 =
\langle\phi_0|V(\alpha)|\phi(\alpha)\rangle$.  We then insert for
$E(\alpha)$, $V(\alpha)$, and $|\phi(\alpha)\rangle$ the series
\eqref{app:V}--\eqref{app:phi(alpha)} and compare coefficients to
obtain the $k$th-order energy shift
\begin{equation}
\label{app:iterE}
E_k = \sum_{k'=1}^{k} \langle\phi_0|V_{k'}|\phi_{k-k'}\rangle .
\end{equation}
It depends on the still unknown correction of the eigenvector,
$|\phi_{k-k'}\rangle$, which we determine from
Eq.~\eqref{app:solution}.  We again insert
Eqs.~\eqref{app:V}--\eqref{app:phi(alpha)} and compare coefficients to
find
\begin{equation}
\label{app:iterPhi}
|\phi_k\rangle
= \frac{Q}{H_0-E_0} \Big\{
      \sum_{k'=1}^{k-1} E_{k'}|\phi_{k-k'}\rangle
     -\sum_{k'=1}^k V_{k'}|\phi_{k-k'}\rangle
  \Big\} .
\end{equation}
Equations \eqref{app:iterE} and \eqref{app:iterPhi} allow the
recursive computation of the series \eqref{app:E(alpha)} for the
eigenvalue shift $E(\alpha)$.

%-----------------------------------------------------------------------------
%\bibliographystyle{elsart-num}
%\bibliography{literature}

%-----------------------------------------------------------------------------

\end{document}